# What Do We Mean by "Accessibility Research"?

A Literature Survey of Accessibility Papers in CHI and ASSETS from 1994 to 2019

KELLY MACK

University of Washington, Seattle, Washington, United States, kmack3@uw.edu

EMMA MCDONNELL

University of Washington, Seattle, Washington, United States, ejm249@uw.edu

DHRUV JAIN

University of Washington, Seattle, Washington, United States, djain@uw.edu

LUCY LU WANG

Allen Institute for AI, Seattle, Washington, United States, lucyw@allenai.org

JON E. FROEHLICH

University of Washington, Seattle, Washington, United States, jonf@cs.uw.edu

LEAH FINDLATER

University of Washington, Seattle, Washington, United States, leahkf@uw.edu

Accessibility research has grown substantially in the past few decades, yet there has been no literature review of the field. To understand current and historical trends, we created and analyzed a dataset of accessibility papers appearing at CHI and ASSETS since ASSETS' founding in 1994. We qualitatively coded areas of focus and methodological decisions for the past 10 years (2010-2019, *N*=506 papers), and analyzed paper counts and keywords over the full 26 years (*N*=836 papers). Our findings highlight areas that have received disproportionate attention and those that are underserved—for example, over 43% of papers in the past 10 years are on accessibility for blind and low vision people. We also capture common study characteristics, such as the roles of disabled and nondisabled participants as well as sample sizes (*e.g.,* a median of 13 for participant groups with disabilities and older adults). We close by critically reflecting on gaps in the literature and offering guidance for future work in the field.

## 1 INTRODUCTION

The *ACM Conference on Accessible Computing* (ASSETS) was founded in 1994 to study the accessibility of digital technologies and to innovate new solutions to real-world accessibility problems. Over the almost three intervening decades, computing has evolved from desktop computers to include mobile touchscreen devices, augmented and virtual reality, and the Internet of Things—all of which pose both technological promise as well as challenges for people with disabilities. While originally a niche topic, accessibility has become a critical focus of industry and the HCI research community as a whole. Indeed, "accessibility" was the second-most popular keyword at ACM CHI 2019 [170].

As the accessibility field matures, surveying its current state and historical context allows us to identify research gaps, discover and question norms with the help of empirical evidence, and provide an entry point for newcomers to understand the field. While focused literature reviews have examined specific subsets of accessibility research (*e.g.,* autism [140,156], visual impairment [19,29]), no broader survey exists. In this paper, we review accessibility research appearing at the two most popular venues for accessible computing—the ACM CHI and ASSETS conferences—from 1994 (the inception of ASSETS) to 2019. Our key research questions include: *Who* does accessibility research focus on? *What* are the stated goals, such as increasing digital accessibility or increasing independence for people with disabilities and older adults? *What* research methods are used, including study design decisions such as sample size, study location, and whether participatory methods are used? And, finally, *how* has accessibility research evolved over time?

To answer these questions, we created a dataset of 836 accessibility papers appearing at CHI and ASSETS from 1994-2019, manually coding 506 of those papers from a recent 10-year period (2010-2019) and analyzing temporal trends in paper counts and keywords for the full dataset. The manual coding captures research foci (*i.e.,* communities of focus, issues addressed, contribution types), methodological decisions (*e.g.,* study methods, sample sizes), and the roles of both disabled and nondisabled participants in accessibility studies. Findings show, for example, that accessibility research focuses disproportionately on the needs of blind and low vision (BLV) users, employs a median sample size of 13 for disabled and older adult participant groups, and commonly pairs some research issues with specific communities of focus (*e.g.,* supporting communication for d/Deaf and hard of hearing people but personal informatics and behavior change for autism-focused work). Complementing these findings, paper counts and keyword frequencies in the full 26 years of the dataset demonstrate that accessibility research is increasing, even outpacing the growth of CHI in the past decade. While there has been an expansion in what user communities receive attention from accessibility researchers, the popularity of BLV work is long-standing. Our data allows us to deeply reflect on the accessibility community and its norms as created and preserved by authors and reviewers.

In summary, this paper contributes: (1) a characterization of current trends in accessibility research, including identification of areas that have received disproportionate attention, areas that are underserved, and patterns in how both disabled and nondisabled participants are included in the research; (2) complementary temporal patterns from 1994-2019, particularly in terms of paper counts and communities studied; (3) reflections and recommendations to guide future accessibility research, identifying opportunities for growth and methodological decisions to question; (4) an open-source dataset of 506 accessibility papers from CHI and ASSETS with our applied qualitative codes and the metadata of 836 accessibility papers from the same venues since 1994[1], enabling future meta-analyses by the HCI accessibility community.

## 2  BACKGROUND AND RELATED WORK

Reflecting on the field of accessibility requires a historical understanding of societal, cultural, and political developments. While many countries have disability rights legislation—including international treaties such as the *United Nations Convention on the Rights of Persons with Disabilities* (CRPD), which stipulates equal access to information and communication technologies, including the Internet [87]—we focus specifically on US policy, which helped influence early CHI and ASSETS accessibility research [50,84]; see Lazar *et al.* [88] for a broader policy overview. Below, we provide background on the history of accessibility research as well as current trends.

### 2.1  The History of Accessibility Research

Throughout history, disability has been cast as a medical problem—a defect that must be cured, rehabilitated, or eliminated if a person is to achieve full capacity as a human [136]. The medicalization of disability began to draw ardent criticism in the 1960s through the emerging disability rights movement [38,134], when activists began framing disability *"as a socially and culturally constructed form of societal oppression"* [38]. This redefinition—ultimately termed the "social model"—shifts the focus of disability from the individual to society and, crucially, distinguishes between impairment as a biological or physical condition and disability as a social and environmental construction [44]. Clarifying this distinction, Kasnitz *et al.* [76] state that *"disability exists when people experience discrimination on the basis of perceived functional limitations"*. The disability studies community continues to debate and build on the social model, such as Kafer's [71] political/relational model, which serves as a *"friendly departure"* and seeks to better encompass the lived experience of impairments, such as pain or chronic illness, and to identify the deeply politicized nature of disability.

The social model and the disability rights movement helped catalyze a new socio-political agenda [133]. In 1973, the US Rehabilitation Act was passed, stating that: *"no otherwise qualified individual with a disability [shall be excluded] from the participation in, be denied the benefits of, or be subjected to discrimination [under any program receiving federal assistance]"* (29 U.S.C. 794d. Section 504. 1973). In a 1986 amendment, the Act was expanded to *"develop and establish guidelines for electronic equipment accessibility"* (Section 508). Although lacking specifics, including definitions of "accessibility" and how guidelines would be enforced, Section 508 was a milestone for accessible technology. Highlighting its importance to the CHI community, early accessibility pioneers Richard Ladner and Gregg Vanderheiden, held a "Section 508" panel at CHI'88 to discuss the development of accessibility guidelines and implications for HCI [84].

In 1990, the Americans with Disabilities Act (ADA) was passed, which, critically, recognized the minority status of Americans with disabilities, extending protections beyond the government sector and requiring places of "public accommodation" to provide people with disabilities appropriate aids or services [38]. Although the ADA has been inconsistently applied to computing technology [86], a 2019 US Supreme Court decision [36] signals increasing acceptance of its applicability. Shortly after the ADA's passage, the *Communications of the ACM* dedicated an issue to disability and computing [10], highlighting *"research related to computers and people with disabilities as a valid area of scientific endeavor"* and arguing that *"design for disability"* was important, cost-effective, and broadly beneficial [50]. The editors invited ACM members to join the *"newly revitalized Special Interest Group on Computers and the Physically Handicapped (SIGCAPH)"* and to attend the inaugural ASSETS conference, which took place in 1994.[2]

---

[1] The dataset, codebook, and analysis scripts are available at the following link: https://github.com/makeabilitylab/accessibility-literature-survey.

Almost three decades later, accessibility has become a key area of CHI [170] and ASSETS is in its 22nd year.[3] CHI 2021's subcommittee on "Accessibility and Aging" defines its focus as: *"technology design for or use by people with disabilities including sensory, motor, and cognitive impairments"* and *"for or use by people in the later stages of life"* [4]. Likewise, the ASSETS 2020 conference advertises itself as the *"premier forum for research on the design, evaluation, use, and education related to computing for people with disabilities and older adults"* and describes relevant topics as including but not limited to: *"new enabling technologies, studies of how technologies are used by people with disabilities, explorations of barriers to access, and evaluations of accessibility education methods"* [3].

**2.2 Current Trends in Accessibility Research**

While no surveys exist of accessibility literature as a whole, researchers have examined sub-areas, including visual accessibility [20,28,52], autism [94,120,140], design with older adults [151], children with "special needs" generally [15], and design philosophies for accessible technology [121,158]. Bhowmick and Hazarika [20], for example, conducted a high-level analysis of 3,010 visual accessibility papers[4] retrieved from four scientific databases, identifying CHI and ASSETS as the most common publication venues. In a more detailed examination of quantitative evaluation practices in technology development for people with visual impairments, Brulé *et al.* [28] analyzed 178 papers published at CHI, ASSETS, TOCHI, and TACCESS from 1988-2019. They found that web browsing was the most popular application area (24.2% of papers), followed by education (14.6%) and mobility (14.6%). Most papers (70.2%) included at least one quantitative empirical evaluation, with a more recent trend to incorporate formative studies. Our work extends similar analyses to the full accessibility literature at CHI and ASSETS, comparing trends across different user populations.

Other surveys have taken a critical discourse analysis approach [140,151]. Vines *et al.* [151] analyzed 644 papers focusing on aging published between 2007 and 2012 across 16 HCI venues. They identify four discourses common in HCI—health economics, socialization (*e.g.,* social isolation), homogeneity (*i.e.,* generalization across older adults), and deficits (*i.e.,* declining abilities)—and call on HCI researchers to avoid the biomedicalization of older adults and to consider the diversity of this population. This deficit discourse is particularly relevant to accessibility research, as also argued by Knowles *et al.* [79], who state that conflating aging with accessibility *"perpetuates negative stereotypes of aging and promotes ageism."* Cognizant of this issue, we only include aging-related papers in our dataset if the authors identified the paper using one of our accessibility-related search terms (Section 3). Another example of a critical discourse analysis comes from Spiel *et al.* [140], who reviewed 185 papers on technologies for autistic children. They focused on the purpose of the technologies and issues of power dynamics, identifying six focus areas such as behavior analysis and social skills. They conclude with the critique that there is *"a predominant focus on corrective, othering approaches"* that *"embody and negotiate […] neurotypical expectations."* We complement and extend these conclusions with quantitative analysis of papers focusing on autism regardless of participant age.

---

[2] ASSETS was initially called the "ACM Conference on Assistive Technologies" and later renamed the "ACM SIGACCESS Conference on Computers and Accessibility". SIGCAPH was renamed SIGACCESS (Special Interest Group on Accessible Computing) in 2003.
[3] ASSETS occurred every other year from 1994 to 2004 and every year since, so ASSETS 2020 marks the 21st ASSETS offering.
[4] It appears that their corpus included extended abstracts like posters in addition to full papers.

Finally, although not literature surveys, accessibility researchers have offered a growing number of critical reflections, calling for greater inclusion of disability studies within accessibility research and for empowering disabled scholars [141]. In an early landmark paper, Mankoff *et al.* [99] called for accessibility researchers to adopt disability studies learnings and to increase representation of disabled people in accessibility work. They cautioned that the medical model, which *"designers of assistive technologies may find […] pragmatically useful"* overemphasizes *"fixing"* the individual rather than addressing larger-scale societal issues related to access and oppression. Frauenberger [44] introduced a "*critical realist perspective*" to assistive technology design, incorporating both intrinsic (*i.e.,* individual) and extrinsic (*i.e.,* structural) factors. Since then, the integration of critical disability studies perspectives to accessibility has only strengthened, with papers examining issues of ableism and epistemic violence in accessibility research [164], the need for greater representation of disabled people in the community [141,155], the conflation of aging and disability [79,151], and interdependence as a guiding value of assistive technology [16]. Our survey method touches on issues of representation by analyzing the use of proxies and of participatory and co-design methods, as well as quantifying who is included as participants in accessibility research.

## 3 METHOD

To understand the current state of accessibility research and reflect on changes over the past three decades, we performed a literature survey of the ACM CHI and ASSETS conferences—two of the most popular venues for accessibility research. This review includes a qualitative analysis of a recent 10-year period (2010-2019, *N*=506 papers) and a briefer quantitative analysis of the 26-year period dating back to the first year of ASSETS (1994-2019, *N*=836).

### 3.1 Dataset Creation

We created the dataset in two phases, focusing first on the 2010-2019 period then expanding to the full 26 years. The final dataset includes only short and long technical papers at ASSETS and CHI (e.g., no posters, keynotes, etc.).

#### 3.1.1 A Recent 10-year Period: 2010-2019

To identify accessibility papers from 2010-2019, we queried the ACM Digital Library (DL) between October and December 2019. Because ASSETS is the *SIGACCESS Conference on Accessible Computing,* we considered all ASSETS papers to be, by definition, about accessibility. CHI papers, however, needed to be filtered to only those that were accessibility-focused, which was complicated by intersections with other sub-disciplines of HCI, particularly research with older adults, health, rehabilitation, and education. Not all papers at these intersections take an accessibility framing, whether explicitly or implicitly. For example, in studying ridesharing with older adults, Meurer *et al.* [105] explicitly distance their work from a deficit or disability discourse (see [151]), while in another study, Harrington *et al.* [54] focus on older adults and health but do not discuss technology accessibility or disability. Thus, so as not to misattribute accessibility or disability framings to these and similar papers, we only considered papers where the authors used one of the following common accessibility terms in the title, abstract, or author keywords: *disab-* (*e.g., disability, disabled*), *access-* (*e.g., accessibility, accessible*), *impair-* (*e.g., impairment, impaired*), and *assistive technolog- (e.g., technology, technologies).*[5] We did not use full text search on the CHI proceedings because our early investigation showed that it resulted in substantially more false positives. We also considered including more disability-specific keywords (*e.g.,* blind, deaf), but did not do so because we were concerned that the results would be biased toward those disability types at the expense of emergent areas that we had not predicted.

---

[5] Specifically, our six search terms were *accessibility*, *disability*, *disabled*, *impairment*, *impaired*, or *assistive technology*, but since the ACM DL search functionality stemmed on keywords, it also provided all results matching the same roots (e.g., *access*, *accessible*). Since running our queries, the ACM DL underwent major revisions which may have changed search logic. Therefore, these queries may not be reproducible.

Our initial search retrieved all 291 technical papers (short or long) that appeared in the ASSETS proceedings from 2010-2019 and identified 594 *candidate* CHI papers from that time period.[6] The candidate CHI papers still included many false positives (enhanced by stemming), such as referring to "access" in the context of security research or a "disabled" user interface feature. Thus, two research team members each independently coded disjoint halves of the candidate papers, marking each one as *relevant*, *irrelevant*, or *needs discussion*. In total, 203 were marked as *relevant*, 49 as *needs discussion*, and 342 as *irrelevant*. The researchers then reviewed each other's *irrelevant* codes, disagreeing in 4 of the 342 instances (1.0%). These disagreements and all *needs discussion* papers were resolved through consensus between the two coders and for a small set (*N*=9), with the entire research team. An additional four *relevant* papers were later classified as *irrelevant* during the more thorough qualitative coding, resulting in 215 *relevant* CHI papers. Combined with the 291 ASSETS papers, the 2010-2019 dataset thus contains 506 papers.

Finally, because our approach may have missed some accessibility papers at CHI, we performed a false negative check. We randomly selected 100 of the 4,927 CHI papers that were *not* among our candidates, and two coders independently reviewed them. Only one paper was arguably an accessibility paper: [125] focused on designing for older adults with memory impairment, but only used our keywords in the full text and not in the metadata fields.

*3.1.2  Expanding to the Full 26 Years: 1994-2019*

We subsequently built on the 506 papers above by adding all 285 ASSETS papers and 45 CHI accessibility papers that appeared in 1994-2009, for a total of 836 papers over the 26-year period. For this expansion, we acquired a metadata file directly from the ACM that included CHI and ASSETS proceedings through January 2020, with each publication's *title*, *authors*, *author keywords*, *venue*, *publication year*, *DOI number*, and *abstract*. The ACM also separately provided a list of the official paper counts per conference per year, with the exception of the first four years of ASSETS. However, we encountered several issues with data quality. First, the aggregate counts of papers from the metadata file did not always match the expected counts per year from their official list. Thus, one author manually reviewed the ASSETS and CHI proceedings frontmatter and/or conference website content (if available) in the full time period (1994-2019), fixing discrepancies between the ACM's official reporting of paper counts and the actual paper counts (*e.g.,* the ACM count erroneously missed short papers for CHI 2008) and removing any papers from the dataset that were not short or long technical papers (*e.g.,* early CHI design briefs). This manual review resulted in a list of 7,209 CHI and ASSETS papers, which was off from what we expected by only six papers (i.e., a 0.1% discrepancy). Second, some of the paper entries had malformed abstracts or missing keywords. We programmatically fixed 271 of the missing abstracts by extracting them from full-text metadata fields, leaving 98 papers (1.4%) without abstracts, and manually retrieved or confirmed the absence of keywords for 131 papers, leaving 9 papers (0.1%) without keyword information.[7]

---

[6] Advanced querying for strings in the title field in ACM DL searches for the string in both the official title and the subtitle.

[7] In our final dataset, 8 CHI and 18 ASSETS papers did not have author keywords, which we verified by looking at the paper PDF. We were unable to gain access to the PDF of 9 ASSETS papers published in the proceedings of the 1994 conference, and their keyword information remains unknown.

We then identified candidate papers from CHI 1994-2009 by string matching on the following terms in the title, subtitle (separate from title in the ACM metadata), abstract, and author keywords: *accessibility*, *assistive tech-*, *disab-*, and *impair-*. This string matching implicitly allowed for some stemming (*e.g., disab* matches to *disability* and *disabled*), but was stricter than the ACM DL search stemming we had used in Section 3.1.1. The result was 74 candidate CHI papers. The same two researchers as in Section 3.1.1 independently checked these candidates to eliminate false positives: three disagreements arose (4.1%) and seven papers (9.5%) were marked as *needs discussion*—all were resolved through consensus. Ultimately, 45 relevant CHI papers remained, which resulted in a 26-year accessibility dataset with 836 papers in total: 260 from CHI and 576 from ASSETS.

Finally, to validate this string-matching method used on the 1994-2009 dataset, we applied it to the 2010-2019 CHI papers. The search identified 355 candidate CHI papers, including *all* 215 that we had manually deemed relevant (see 3.1.1), plus 140 false positives that we had eliminated. We again also checked for false negatives in the 1994-2009 data by randomly selecting 100 papers from CHI 1994-2009, finding that two of them did not use our keywords but could arguably appear at ASSETS [9,130]. Thus, we acknowledge that a small set of papers may not be included in our dataset, which was a tradeoff we considered necessary to eliminate false positives.

**3.2 Analysis**

We qualitatively coded the 506 papers from 2010-2019, and programmatically analyzed all 26 years of data (N=836). We note that the process of creating and analyzing the dataset involved subjective judgments, and that this research was conducted by white and Asian scholars, two of whom identify as disabled. Authors ranged from first year graduate students to professors who have been publishing accessibility research in CHI and ASSETS for about 15 years. We come to this analysis with deep personal and academic commitments to accessibility and acknowledge that our scholarship reflects our own biases and beliefs.

*3.2.1 Qualitative Coding of a 10-year Snapshot*

To analyze the 506 papers from 2010-2019, we used an iterative process to develop and apply a codebook. The initial codebook included deductive codes based on our research questions, such as user communities of focus, technologies, and study methods. Three authors then went through three iterations of independently applying and updating the codebook, using a randomly selected set of 25 papers for each iteration. After each set, the coders came together to compute interrater reliability (IRR) using Krippendorff's alpha [169], refine or eliminate existing codes, add new inductive codes, manually resolve disagreements through consensus, and discuss the outcome with the full research team. The final codebook included 10 overarching categories with 2-10 subcodes each. Excluding "Other" subcodes, IRR across subcodes ranged from .63 to .91 and was on average 0.76 (*SD*=.09)[8]. Finally, the remaining papers were split into approximately three sets and coded separately by the same three coders who developed the codebook. Halfway through their sets, all three authors coded an overlapping set of five papers to provide an opportunity for discussing any emergent concerns; disagreements on this small set were resolved through consensus.

The final codebook is summarized in Table 1. While full code definitions can be found in the Supplementary Materials, we offer a few notes here. *Community of focus* is the accessibility-related population or community being studied or positioned by the authors as benefiting from the research. For *Participatory design use*, we coded "yes" if a paper explicitly stated that they used "codesign", "co-design", or "participatory design" in their study method; there were a few papers that used phrases like "participatory approach", which we coded as "no" (*e.g.,* [49,98]). *Use of proxies* refers to when someone other than the target user is asked to speak to the thoughts, preferences, or behaviors of a person with a disability in place of that person themselves [7,112], which is distinct from participants who were acting in strictly a stakeholder or caregiver role. *Issue addressed* examined the research goal of each paper, while *Contribution type* identified the paper's research contribution(s) according to Wobbrock and Kientz's *Research Contributions in Human-Computer Interaction* [159]. Finally, *Participant count* captured the number of participants in each participant group in the user study or studies in the paper. For these counts, we grouped participants into one of 12 categories covering the *Participant groups* codes, but also more specifically breaking down *people with disabilities* into one of the *Communities of focus* codes. We additionally extracted papers about crowdworkers from the "other" code, since their sample sizes were considerably larger from the other papers coded as "other".

---

[8] We did this IRR exercise as we were establishing our final dataset. As such, in our second and third round of coding, we coded one paper in each round that did not qualify for the final dataset (two ASSETS experience reports).

*3.2.2 Programmatic Analysis of 26-year Temporal Trends*

To complement the qualitative coding, we programmatically examined paper counts and keyword frequencies over the full 26-year period (*N*=836 papers). For keyword analysis, two researchers manually reviewed all of the 2,101 unique author keywords appearing at least twice in the dataset (*N*=511 keywords), categorizing each one as being thematically related to users (i.e., community of focus), technology, method, or none of the above. Then, within each of these three themes (user, technology, and method), the researchers grouped highly similar keywords into higher-level categories informed by our qualitative codebook; for example, the user-related keywords of "mobility", "motor disabilities" and "motor impairments" were combined into a "motor/physical" category. Because author keywords tended to be sparse, we then enriched the keyword data by searching for additional occurrences of all 2,101 author keywords in paper titles and abstracts, looking for exact token matches. This enrichment process allowed us to identify user, technology, and method keywords for an additional 202, 203, and 398 papers, respectively. However, ultimately, 125 papers (15.0%) had no user keywords, 108 papers (12.9%) had no technology keywords, and 285 papers (34.1%) had no method keywords.

Table 1: The final codebook with 10 code categories and 52 subcodes; the average IRR is computed across the subcodes for each code. When applicable, the IRR calculations do not include the code of "other". Multiple refers to "multiple codes can apply."

| Category | Codes | Mean IRR | Multiple? |
|---|---|---|---|
| Community of focus | Blind or low-vision (BLV); deaf or hard of hearing (DHH); autism; intellectual or developmental disability (IDD); motor or physical impairment; cognitive impairment; older adult; general disability or accessibility; other | 0.91 (*SD*=0.17) | Yes |
| Study method | Controlled experiment; interview; survey; usability testing; case study; focus group; field study; workshop or design session(s); other | 0.75 (*SD*=0.32) | Yes |
| Participatory design use | Yes; no | 0.74 (*SD*=0.00) | No |
| User study location | Near/at a researcher's lab; neutral location; online or remote; unclear location; participants' residence, school, work, or similar location; or no user study; other | 0.71 (*SD*=0.20) | Yes |
| Participant groups | People with disabilities; older adults; caregivers; specialists (*e.g.,* therapists or teachers); people without disabilities; no user study; other | 0.88 (*SD*=0.20) | Yes |
| Use of proxies | Yes; no | 0.68 (*SD*=0.00) | No |
| Ability-based comparisons | Yes; no | 0.80 (*SD*=0.00) | No |
| Issue addressed | Increasing independence; increasing digital access; increasing physical access; increasing understanding of users; supporting communication; or personal informatics and changing behavior; other | 0.63 (*SD*=0.34) | Yes |
| Contribution type | Empirical; artifact; methodological; theoretical; dataset; survey | 0.76 (*SD*=0.39) | Yes |
| Participant count | Blind or low-vision; deaf or hard of hearing; autism; intellectual or developmental disability; motor or physical impairment; cognitive impairment; older adult; general disability or accessibility; caregivers; specialists (*e.g.,* therapists or teachers); people without disabilities; other-crowdworkers; other | N/A | Yes |

## 4 RESULTS

We first characterize the current state of accessibility research (2010-2019) in terms of focus areas, study methods, and inclusion of disabled and nondisabled participants before complementing these findings with a programmatic analysis of temporal trends from the full 26-year period (1994-2019).

### 4.1 Communities of Focus, Research Problems and Contributions

To understand where the accessibility research field invests its effort and what gaps might exist, we summarize what communities and research problems receive attention, and the types of research contributions made.

#### 4.1.1 Communities of Focus

As shown in Table 2, accessibility research focuses disproportionately on visual accessibility: almost half of the papers (43.5%, *N*=220/506) were aimed at addressing the needs of BLV people followed by a precipitous drop to people with motor or physical disabilities (14.2%, *N*=72) and people who are deaf or hard of hearing (11.3%, *N*=57). The remaining communities of focus each accounted for under 10% of papers, including people with cognitive impairments (9.1%, *N*=46), older adults (8.9%, *N*=45), autism (6.1%, *N*=31), and IDD (2.8%, *N*=14). Additionally, 9.1% (*N*=46) of papers aimed to address the disability community in general, including investigations of assistive technology [135], interviews with disability activists [91], and accessibility adherence/knowledge across professions [106,123]. The code "other" was also applied to 9.1% (*N*=46) of papers, of which 20 had no additional code; further analysis revealed that these 20 focused primarily on color vision accessibility (*N*=7), such as work by Flatla *et al.* [41,96], in addition to mental health [127,163], special education [89], and learning disability topics more generally [25,126].

Table 2: The frequency of applied codes for *community of focus*, *issues addressed*, and *contribution type*.

| Community of Focus | Papers w/ Code | This Code Only | Issue Addressed | Papers w/ Code | This Code Only | Contribution Type | Papers w/ Code | This Code Only |
|---|---|---|---|---|---|---|---|---|
| BLV | 220 (43.5%) | 208 (41.1%) | Digital Access | 186 (36.8%) | 122 (24.1%) | Empirical | 305 (60.3%) | 171 (33.8%) |
| Motor/Physical | 72 (14.2%) | 59 (11.7%) | Understanding Users | 139 (27.5%) | 89 (17.6%) | Artifact | 281 (55.5%) | 182 (36.0%) |
| DHH | 57 (11.3%) | 43 (8.5%) | Physical Access | 105 (20.8%) | 26 (5.1%) | Theoretical | 44 (8.7%) | 6 (1.2%) |

| | | | | | | |
|---|---|---|---|---|---|---|
| Cognitive | 46 (9.1%) | 29 (5.7%) | Independence | 93 (18.4%) | 14 (2.8%) | Methodological 16 (3.2%) 2 (0.4%) |
| General Disability | 46 (9.1%) | 31 (6.1%) | Communication | 81 (16.0%) | 45 (8.9%) | Dataset 7 (1.4%) 2 (0.4%) |
| Older Adult | 45 (8.9%) | 29 (5.7%) | Behavior Change | 39 (7.7%) | 19 (3.8%) | Survey 3 (0.6%) 0 (0.0%) |
| Autism | 31 (6.1%) | 21 (4.2%) | *Other* | 59 (11.7%) | 25 (4.9%) | |
| IDD | 14 (2.8%) | 8 (1.6%) | | | | |
| *Other* | 46 (9.1%) | 20 (4.0%) | | | | |

Because papers could be coded with more than one community of focus, 7.1% (*N*=36) received multiple codes (not considering the code "other"). We analyzed co-occurrence patterns and found that some communities were much more likely to appear on their own than with others. Most notably, a vast majority of BLV papers (94.5% of 220 papers) had a singular focus. Of the twelve exceptions, eight were marked with general disability, two with older adult, two with motor/physical, and one with cognitive impairment. In contrast, the IDD code was the most likely to co-occur with another code (*N*=6, 42.9% of 14 papers), which is often due to its appearance with people with autism (*N*=5, 35.7%) and people with cognitive impairments (*N*=1, 7.1%). The older adults code also commonly co-occurred (*N*=16, 35.6% of 45 papers) which emphasizes the common disability-related focus of research with older adults in the accessibility community, such as designing for older adults with tremor [113,153] or mild cognitive impairment [97,103].

*4.1.2 Issues Addressed*

When examining the research aims of papers in our dataset, the three most common objectives were to increase digital access through a technology innovation (36.8%, *N*=186/506), understand user needs, preferences and abilities (27.5%, *N*=139), and increase physical world access (20.8%, *N*=105)—see Table 2. We also identified and coded for three more specific and less frequent areas of investigation: increasing independence (18.4%, *N*=93), supporting communication (16.0%, *N*=81), and personal informatics and/or behavior change (7.7%, *N*=39); an additional 4.9% of papers (*N*=25) received the sole code of "other". Again, multiple codes could apply, such as if a paper aimed to increase physical world access but also framed that innovation as increasing independence (*e.g.,* [18,61]).

Perhaps more interestingly, research papers addressed different issues depending on the community of focus. The DHH community papers contrasted most strongly with overall trends: almost two thirds of DHH papers (64.9% of 57 total) addressed communication, such as real-time captioning [83,119] or signing avatar technology [65,78]. Similarly, papers focusing on the autism community most commonly addressed behavior change and increasing understanding of users—each of which appeared in 32.3% of the 31 autism papers—while increasing independence (42.9%) and behavior change (35.7%) were the top two issues for the 14 IDD papers. For example, a smartwatch application for people with IDDs aimed to change behavior by *"notifying the student to keep [himself]/herself focused, to ask or answer questions, to participate in group discussions and to moderate his/her voice (speaking more or less loudly)"* [166], while for autistic students: *"MOSOCO suggests skill appropriate do's (e.g., "smile if somebody looks at you") and don'ts (e.g., "don't stare")"* [39]. In contrast, the remaining communities of focus more closely mirrored overall trends. For BLV and motor/physical papers, increasing digital access was the most common issue addressed (48.2% of 220 BLV papers and 40.3% of 72 motor/physical papers), while for the older adult, cognitive impairment, and general disability communities, the most common issue addressed was to understand users (46.7% of 45, 37.0% of 46, and 47.8% of 46 papers, respectively).

*4.1.3 Contribution Type*

The above issues were addressed primarily through empirical contributions (60.3%, *N*=305/506)—often to understand a populations' view toward or use of a technology (*e.g.,* [27,75,78])—or artifact contributions (55.5%, *N*=281), building a tool and evaluating it with users (*e.g.* [69,101,131]). These two contribution types frequently occurred in combination (18.4%, *N*=93). The remaining five contribution types occurred considerably less frequently, with theoretical being the next most popular (8.7%, *N*=44), including Mankoff *et al.'s* disability studies as a source of critical inquiry paper [100] and Bennett *et al.'s* work on interdependence as a frame for assistive technology [16]. Only 16 papers (3.2%) made methodological contributions, such as Trewin *et al.'s* [147] *Usage of Subjective Scales in Accessibility Research* and Holone and Herstad's [62] *Three Tensions in Participatory Design for Inclusion*. Dataset contributions were even rarer, at 1.4% (*N*=7) of papers [12,42,66,90,150,160,162]—two of which received no other contribution type code: Flores and Manduchi's *WeAllWalk* dataset [42] and Wolter *et al.'s CADENCE* corpus for inclusive voice interface design [160]. Finally, literature survey was the least common contribution type, occurring in only three papers (0.6%), including Carter *et al.'s* survey of autism research [33], Abbott *et al.'s* work on anonymization practices in aging and accessibility research [1], and the aforementioned Trewin *et al.*'s [147] work on subjective scales.

**4.2 Prevailing Research Methods**

While the above section analyzed where the accessibility community invests attention, here we explore *how* this research was conducted, *who* is included in research studies, and other decisions related to study design.

*4.2.1 What overall research methods are most popular?*

Reflecting the human-centered design orientations of the ASSETS and CHI communities, accessibility papers overwhelmingly include user studies: 94.3% (*N*=477) of 506 papers. For the 29 papers that did *not* include user studies, research methods included analyses of existing content (*e.g.,* forum posts [30,72]), algorithmic/system analyses (*e.g.,* [35,96]), literature reviews (*e.g.,* [1,147]), and theoretical contributions (*e.g.,* [51,124]).

Focusing on only the 477 user-study papers, the three most common methods—accounting for 84.1% of papers (*N*=401/477)—were interviews (42.1%, *N*=201), usability testing (41.7%, *N*=199), and controlled experiments (34.6%, *N*=165). Interestingly, a majority of papers (56.4%, *N*=269) were coded with multiple user study methods, with surveys, interviews, and focus groups most commonly co-occurring with others. For the papers that included only a single method (35.6%, *N*=180/506), the most common methods were controlled experiment (30.6%, *N*=55/180) and usability testing (25.6%, *N*=46/180). Additionally, 16.1% of user study papers (*N*=77/477) were coded as other. For these, besides four papers [42,43,77,146] that evaluated algorithm/model performance, all remaining other codes added nuance to a non-other code (*e.g.,* specifying that usability testing was performed via contextual inquiry [6,23]).

Table 3: The frequency of applied codes for *study method*, *study location*, and *participant group* for the 477 user-study papers.

| Study Method | Papers w/ Code | This Code Only | Study Location | Papers w/ Code | This Code Only | Participant Group | Papers w/ Code | This Code Only |
|---|---|---|---|---|---|---|---|---|
| Interviews | 201 (42.1%) | 27 (5.7%) | Unclear | 189 (39.6%) | 133 (27.9%) | People w/Disability | 404 (84.7%) | 214 (44.9%) |
| Usability Testing | 199 (41.7%) | 46 (9.6%) | Home/Freq. Loc. | 138 (28.9%) | 85 (17.8%) | People w/out Disab. | 110 (23.1%) | 5 (1.0%) |
| Controlled Exp. | 165 (34.6%) | 55 (11.5%) | Lab | 130 (27.3%) | 93 (19.5%) | Specialist | 81 (17.0%) | 9 (1.9%) |
| Survey | 122 (25.6%) | 6 (1.3%) | Online/Remote | 98 (20.5%) | 48 (10.1%) | Caregivers | 45 (9.4%) | 4 (0.8%) |
| Workshop/Design | 88 (18.4%) | 15 (3.1%) | Neutral Loc. | 32 (6.7%) | 15 (3.1%) | Older Adult | 40 (8.4%) | 15 (3.1%) |
| Field Study | 85 (17.8%) | 22 (4.6%) | *Other* | 7 (1.5%) | 1 (0.2%) | *Other* | 53 (11.1%) | 23 (4.8%) |
| Focus Groups | 28 (5.9%) | 4 (0.8%) | | | | | | |
| Case Study | 19 (4.0%) | 1 (0.2%) | | | | | | |
| *Other* | 77 (16.1%) | 4 (0.8%) | | | | | | |

*4.2.2 Where are studies conducted?*

HCI user studies are typically conducted in research labs, but allowing for other locations—such as from home or in community centers—may broaden who is able to participate. Of the 477 papers with user studies, 405 (84.9%) specifically mentioned a location for at least one study, though nearly 40% still had at least one study location that was unclear (39.6%, *N*=189/477). While a substantial portion were laboratory/on-campus sites (32.1%, *N*=130/405), slightly more were at participants' home, work, or a place they visit frequently (34.1%, *N*=138). Remote participation (24.2%, *N*=98) and neutral locations like a community library (7.9%, *N*=32) were also common. Non-campus locations were often selected to improve aspects of the research, like increasing ecological validity by studying a behavior or social interaction in context (*e.g.,* [55,114,157]) or allowing for easier recruitment of communities (*e.g.,* summer camp [2], care facility [152], or school [47]). Other times, locations were chosen to reduce travel burden or increase accessibility (*e.g.,* [26,110,143,168]); for example, Zou *et al.* explained, "*Participants were encouraged to choose a location of their preference. Fifteen chose their home, one chose a public library, one chose his office, and one chose the researcher's home*" [168].

*4.2.3 Who is included in accessibility studies and what are the sample sizes?*

We examined *who* and *how many* participants are included in accessibility studies. For the 477 user-study papers, we found that most (90.1%; *N*=430/477) included participants with disabilities and/or older adults (Table 3). For the 47 papers (9.9%) that did not, almost half included specialists and/or caregivers (38.3%, *N*=18/47) in proxy roles, which we describe further in Section 4.3. The remaining 29 papers either had nondisabled participants (*N*=6) and/or had been marked as "other" (*N*=24). For the former, papers focused on educating nondisabled people on accessibility issues [21,81], running disability simulations [48], and having nondisabled people test technologies that were designed to help people with disabilities [34]. For the 18 papers coded only as "other", 10 focused on ASL signers in general, without requiring that they be d/Deaf or hearing, such as papers evaluating sign language video quality [145].

Recruiting users in accessibility research can be challenging, especially when working with highly specific populations [132]. To assess sample size practices across these participant groups, we analyzed participant numbers in the 477 user-study papers. Twenty-six papers did not clearly report on participant numbers (e.g., [128]), leaving 451 papers for analysis. For this analysis, we further split out crowdsourced participants from the more general "other" code.

Overall, the median number of participants per paper was 18 (*IQR*=29.0, *M*=830.1, *SD*=16,140.3). However, breaking down these numbers into multiple participant groups per paper shows that samples specifically of participants with disabilities or older adults have a median of only 13 (*IQR*=13.0, *M*=62.0, *SD*=478.8). Table 4 shows this detail, with sample sizes ranging from a median of 9 (autism) to 28 (DHH), with many in the 9-16 range.

Distributions were also highly skewed, with a few studies having 1,000+ participants (*e.g.,* [92])—particularly crowdsourced studies—and seven papers overall having only one or two participants. These smallest sample sizes came mostly from field studies [5,22,68,90] such as Alankus *et al.*'s close work with one participant in studying stroke therapy [5]. Caregiver and specialist groups had relatively small sample size medians of 7.5 and 7.0, respectively, reflecting the advisory role these groups often play.

Table 4. Number of participants broken down by type of *Participant Group* and *Community of Focus*, the 451 of 477 papers with user studies and that clearly reported on sample size for at least one participant group.

| Group | Med | Mean | IQR | Range | Total Papers |
| --- | --- | --- | --- | --- | --- |
| *Other*-Crowdworkers | 153 | 27,500.0 | 755.0 | 15-351,960 | 13 |
| DHH | 28 | 44.5 | 41.5 | 1-284 | 35 |
| General DA | 19 | 24.0 | 14.5 | 3-55 | 4 |
| Nondisabled | 16 | 97.8 | 26.0 | 2.0-3,323 | 105 |
| Cognitive | 16 | 40.3 | 33.0 | 3-328 | 29 |
| Older Adult | 15 | 32.8 | 18.3 | 2-519 | 38 |
| BLV | 13 | 88.8 | 12.0 | 1-7,398 | 208 |
| IDD | 11.5 | 16.4 | 10.3 | 2-48 | 10 |
| *Other*/Unclear | 11 | 47.3 | 22.0 | 1-566 | 61 |
| Motor/Physical | 10 | 15.3 | 11.0 | 1-102 | 61 |
| Autism | 9 | 12.9 | 5.0 | 2-75 | 23 |
| Caregivers | 7.5 | 11.0 | 8.8 | 1-43 | 34 |
| Specialists | 7 | 14.2 | 11.0 | 1-80 | 69 |

*4.2.4 How widely are participatory methods used?*

Participatory design (PD) provides a method and vision for involving users of technology directly in its design [45]. Though PD has a long history within HCI [111], there are ongoing debates about what constitutes PD and the extent to which it *"has become too diluted"* [13], with papers using the term "participatory" for a wide variety of design activities and study lengths. For accessibility research, in particular, there are unique tensions, including the additional burden placed on PD participants for their "access labor" [17] and the inaccessibility of traditional participatory methods to certain communities, such as autistic people [46] or people with aphasia [73].

In our analysis, we found that 10.3% (*N*=49/477) of papers with a user study identified themselves as using PD or co-design. Some studies engaged participants in rich design processes such as crafting and bodystorming [104], play-acting [73], and ideating on prompts [64,67]. Others used participatory design to test an evolving research prototype, with the same or different users [53,115,137]. The number of sessions varied from single instances [64,115] to as many as 20 sessions [161]. Almost all participatory-design papers included disabled participants and/or older adults (95.9%, *N*=47/49), and a substantial portion also included other stakeholders, such as specialists, therapists, and teachers (40.8%, *N*=20) and/or caregivers (20.4%, *N*=10). Wu *et al.* [161], for example, included caregivers and specialists in the participatory design of a calendar app for memory loss, and Waddington *et al.* [154] involved both therapists and BLV people in designing an app for therapeutic exercises. Only two papers employing participatory design did *not* include disabled participants or older adults, both of which focused on enabling nondisabled stakeholders (teachers, occupational therapists) in 3D printing technologies for people with disabilities [31,59]. Finally, while PD was used across all communities of focus, it was most prevalent in papers on cognitive impairment (21.7%, N=10/46) or IDD (21.4%, *N*=3/14) compared with 6-17% for other communities of focus.

### 4.3 The Role of Nondisabled Participants, Caregivers and Specialists

Accessibility work typically centers people with disabilities, yet caregivers, specialists, and nondisabled participants are also involved, often as stakeholders but also as *proxies* or *comparison groups*—two roles for which we explicitly coded.

*4.3.1 How and when are proxies used?*

Proxy use occurred in 8.0% of all user-study papers (*N*=38/477), most often in conjunction with participants with disabilities (*N*=32 of the 38 papers). Papers did not typically specify rationale for *why* proxies were used; however, for those that did, reasons ranged from lessening the burden on participants [57] to working around communication difficulties [109]. As an example of the latter, "*Since some of our participants had verbal difficulties, we asked staff members to identify the preferred interests for each participant prior to the date of the study*" [109].

Typically, proxies were specialists (65.8%, *N*=25/38) and/or caregivers (44.7%, *N*=17) and were asked to offer a disabled individual's perspective of a situation [24,85,138], behavior [8,165], and/or new technology [32,116]. For example, in a study from Boyd *et al.* [24], a parent discussed her autistic child's behavior: "*I feel he doesn't know how close to get or how to close not to get [to other people]. It's like an abstract idea for him*". Similarly, in a survey designed for people with motor impairments, Anthony *et al.* [8] reported that "*in many cases, a third person completed the survey on the user's behalf, either because the primary user was a child or because it was difficult for the primary user to do*". This input from proxies often complemented data from disabled participants, such as Carrington *et al.*'s [32] use of focus groups with doctors, therapists, and family members in addition to sessions with power wheelchair users.

When examining the use of proxies as a function of community of focus, we found that proxies were most common in studies focusing on autism (22.6%, *N*=7/31), followed by cognitive impairment (17.4%, *N*=8/46) and IDD (21.4%, *N*=3/14). Common reasons included seeking the domain expertise of specialists or caregivers (*e.g.,* IDD [107,144], cognitive impairment [139]) and communication difficulties (*e.g.,* autism [53,109], cognitive impairment [85]). At the other end of the spectrum, there were no instances of DHH-focused papers that included proxies.

*4.3.2 How and when are disabled and nondisabled participants compared?*

We identified 65 papers (13.6%) that compared disabled and nondisabled participants (*e.g.*, blind *vs.* sighted users or people with and without motor impairments), often to understand people's needs, experiences, or perceptions across varied abilities [11,82,108,126,149] but also to establish a comparative "baseline" or "control group" (*e.g.,* [60,92,102,167]). As examples of the former, Kushalnagar *et al.* [82] studied how caption needs differ depending on hearing ability in a classroom deployment and Oh *et al.* examined differences in touchscreen use between BLV and sighted users as reported in an online survey [118]. These papers most commonly included controlled experiments (*N*=41, 63.1%), with nondisabled participants often acting as a control group. For example, Findlater *et al.* [40] investigated whether touch screen interaction "*reduce[d] the performance gap between older and younger adults*" compared to using a mouse.

Another use of ability-based comparisons was to understand differences between ability level to allow for collaboration between disabled and nondisabled individuals. Kacorri *et al.* [70], for example, compared images taken by blind and sighted users to "*explore the benefits of having a sighted person training a blind user's personal object recognizer.*" Comparison can also be used to create or evaluate equitable experiences across ability levels, such as da Rocha Tomé Filho *et al.*'s [129] assessment of BLV and sighted participants' experiences in playing adapted board games.

**4.4 Historical Context: Programmatic Analysis of 26-year Temporal Trends**

While the qualitative coding analysis revealed current foci and methods employed in accessibility work, here we turn to longer-term temporal trends via a programmatic analysis of overall paper counts and keywords appearing in the author keywords, titles, and abstracts of papers from 1994-2019.

*4.4.1 The growth of accessibility research*

As shown in Figure 1, accessibility research has grown rapidly over the past 26 years: from 22 papers in 1994 (0 at CHI and 22 at ASSETS) to 95 in 2019 (55 at CHI and 40 at ASSETS). This growth is particularly pronounced within the last five years at CHI, with the number of accessibility-related CHI papers overtaking the overall number of ASSETS papers in 2018, and the growth in accessibility papers outpacing the growth of CHI itself (Figure 1, right). In 2019, accessibility papers made up 7.8% (55 of 702 papers) of all CHI papers—and the keyword "accessibility" was the second most common overall (behind "virtual reality") [170].

*4.4.2 Temporal trends in communities of focus, technologies, and methods*

As described in Section 3.2.2, to analyze more specific temporal trends over the 26 years, we categorized author keywords as relating to one of three themes (if applicable): communities of focus (*i.e.,* users), technologies, and research methods. Within each theme, we combined similar keywords into higher-level keyword groups, such as grouping "low vision" and "blind" into "BLV"; see Supplementary Materials for the full list of groupings. We then counted the number of papers per year that used each of those keyword groups, normalizing the count if a given paper employed more than one keyword group within a theme. For example, if a paper had both BLV and older adult keywords, the count would be 0.5 toward BLV and 0.5 toward older adults. Table 5 shows the top 12 keyword groups for each of the three themes over the full 26 years, in addition to proportions of papers using those keywords over time (sparkline plots). To smooth the sparkline plots, we aggregated papers into five bins: 1994-1999, 2000-2004, 2005-2009, 2010-2014, and 2015-2019.

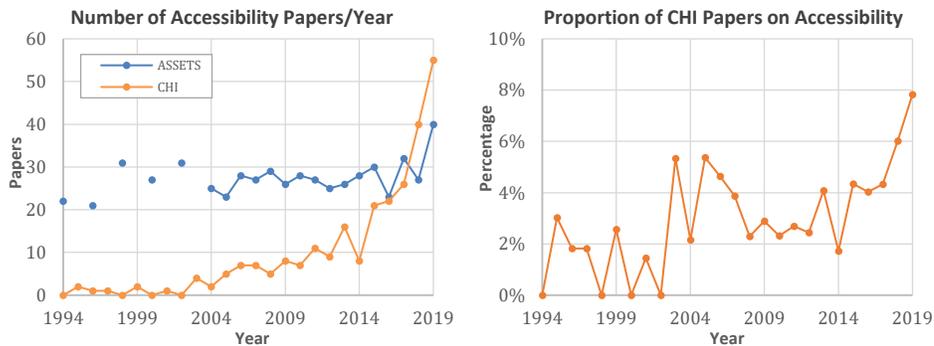

Figure 1: Accessibility paper counts at CHI and ASSETS over time (*left*) show that accessibility is growing as a field, especially the past five years. The percent of CHI papers on accessibility (*right*) shows that accessibility research has grown as a proportion of all CHI papers, reaching nearly 8% of CHI papers in 2019.

The main trends for community of focus are largely consistent with the earlier qualitative analysis, showing the high proportion of attention to the BLV community followed by motor/physical and DHH communities. Papers on older adults and people with cognitive impairments also appear consistently since the earliest years, with other communities of focus only appearing more recently, including autism and neurodiversity (2000-2004), color vision deficiency (2005-2009), and mental health (2010-2015). "Children" is a relatively common keyword for which we did not explicitly code, but that reflects that accessibility work often focuses on children's specific needs (e.g., [47,53,55]). Another new keyword here is "literacy", which can relate to areas ranging from tactile media consumption [142] to literacy issues related to understanding terms of service information [95].

For technology keywords, which we had not reported on in the qualitative analysis, temporal trends follow broader technology trends over time. "Web" is the all-time most popular technology keyword, but it dropped off in the past 10 years. "Mobile", in contrast, grew substantially before plateauing in the last five years. The general term "user interfaces" has also dropped since the early years, perhaps because it is redundant with HCI. To understand more recent changes in technology foci, we also identified which popular technology keyword groups have only appeared in the last 10 years. Examining the top 20 keyword groups from 2010-2019 reveals six that were not present in the top-20 list from 1994-2009: games, wearable computing, social computing, 3d printing and DIY, AR/MR/VR (augmented, mixed or virtual reality), and collaboration tools—all representing recent technology trends in accessibility and HCI more broadly.

Table 5. All keyword groups for community of focus (excluding "other/unclear", which accounted for 4.5% of papers with community of focus keywords), plus the top 12 most popular technology and method keywords over the full time span of 1994-2019. Paper counts are normalized as described in the main text. The sparklines show the proportion of papers that use at least one keyword from a given group over time, with each dot representing a time chunk: 1994-1999, 2000-2004, 2005-2009, 2010-2014, and 2015-2019.

| Community of Focus | | | Technology | | | Method | | |
|---|---|---|---|---|---|---|---|---|
| Keyword Group | Papers (*N*) | Proportion over Time | Keyword Group | Papers (*N*) | Proportion over Time | Keyword Group | Papers (*N*) | Proportion over Time |
| BLV | 300.0 | | Web | 68.1 | | Design | 214.5 | |
| Motor / physical | 89.4 | | Mobile | 47.6 | | Evaluation | 90.7 | |
| DHH | 71.8 | | Interaction | 37.3 | | User study | 67.4 | |
| Older adults | 50.6 | | Auditory | 36.2 | | Usability | 49.1 | |
| Cognitive impairment | 45.5 | | Input: general | 29.5 | | Interviews | 38.3 | |
| Children | 40.7 | | User interfaces | 26.1 | | Participatory / co-design | 15.2 | |
| General | 31.0 | | Speech/voice | 25.1 | | Crowd-sourcing | 12.3 | |
| Autism and neurodiversity | 23.4 | | Video | 22.9 | | Scenarios | 10.0 | |
| Literacy | 9.2 | | Input: pointing | 22.8 | | Case study | 9.3 | |

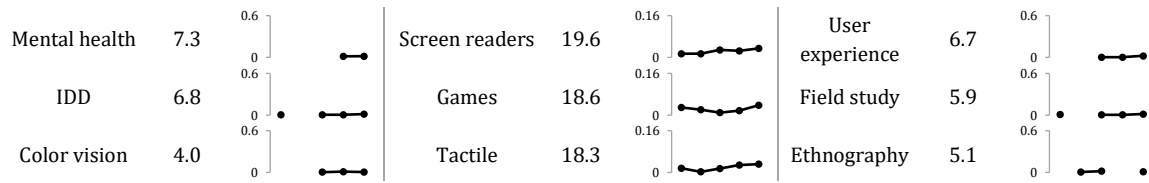

| | | | | | | | | | |
|---|---|---|---|---|---|---|---|---|---|
| Mental health | 7.3 | | | Screen readers | 19.6 | | User experience | 6.7 | |
| IDD | 6.8 | | | Games | 18.6 | | Field study | 5.9 | |
| Color vision | 4.0 | | | Tactile | 18.3 | | Ethnography | 5.1 | |

Finally, for methods, the generic terms "design", "evaluation", and "user study" are the most popular. Perhaps reflecting larger language and conceptual changes in HCI research and practice, "usability" drops over time while "user experience" only began to appear in 2005-2010. In terms of other specific user study methods, interviews appear to be increasing over time, only appearing for the first time in 2000-2004 and growing to 10.9% of method keywords in 2015-2019. However, we note that this number is considerably lower than the 42.1% of papers that we had coded as employing an interview method in Section 4.2.1, which may reflect the relative lack of method keywords compared to community of focus and technology keywords (34.1% of papers had no method keywords). Other methods are infrequent but consistently present over the years, in line with our qualitative coding: participatory design, case studies, field studies, and the related method of ethnography. Finally, crowdsourcing unsurprisingly appears only within the last 10 years.

## 5 DISCUSSION

Accessibility research has grown substantially over the past few decades, proportionally outpacing the growth of CHI itself. This is a moment to celebrate and to reflect. In this first broad literature survey of accessibility work, we describe a sprawling research field that, while disproportionately focusing on blind and low vision users, encompasses many other communities and addresses a range of issues. We surface norms in accessibility research—norms that have been shaped by accessibility researchers, reviewers, and journal and conference leadership and that all members of our community should continually question and reassess. To help guide future accessibility research, we recap and discuss the more critical gaps and trends from our analysis, as well as implications for conducting similar literature surveys.

### 5.1 Current Foci and Growth Opportunities in Accessibility Research

Underpinning our work is the question of *"what is accessibility research?"*. Accessibility has a home within HCI but touches on computer science, disability studies, gerontechnology, health, rehabilitation engineering, and other areas. In focusing on CHI and ASSETS, the two most popular conferences for HCI-focused accessibility work, we describe the core field but do not fully represent these peripheries nor do we fully capture contributions that might be more common in journals than in conferences (e.g., literature surveys or multi-study papers). Our decision to filter CHI papers to those that self-identified as accessibility focused [79] also further shaped our dataset and the conclusions we can draw, defining an expansive set of "accessibility" papers but also undoubtedly missing some that could arguably qualify.

Within that context, our analysis reveals *who* and *what* research problems are valued by the community. In terms of *who*, BLV people are by far the most common community of focus. We suspect that this skew is due to multiple interdependent factors including funding mechanisms, the popularity of BLV people in public disability discourse, and the apparent concreteness of visual accessibility problems to HCI researchers (e.g., how to make computer I/O accessible to a blind user). On the other hand, autism, IDD, cognitive impairment, and other areas are less common or have only appeared in the accessibility literature more recently. These lattermost areas and the equally important communities of focus that were coded as "other" point to opportunities for growth. Mental health issues and chronic illness, for example, are prominent in our society but not well-represented in the dataset, suggesting that HCI researchers could more actively pursue work in this space from a disability perspective. Adopting an accessibility or disability framing introduces questions around how technology and norms can adapt to be more inclusive of these individuals. Relatedly, we found that few papers included multiple communities of focus (7.1%) or people with multiple disabilities (<1.0%)[9]: a critical omission that needs future work. Lastly, though not captured in our study, future work can follow Ogbonnaya-Ogburu *et al.* [117] to explicitly consider how participants' Deaf or disabled identities intersect with other aspects of their identity, such as race and gender, and the implications for design.

Accessibility research is also defined by *what* problems it tackles. Unsurprisingly, digital and physical accessibility problems, and attempts to understand related user needs are common. Perhaps more useful for shaping future work is to question why other research problems are popular only within some user communities. Focusing on communication for DHH users (64.9% of DHH papers), for example, may well be appropriate for that community, who can experience communication as a disabling circumstance in a dominant hearing culture. In contrast, people with autism and IDD have repeatedly voiced that they do not want or need to change their behavior to make others more comfortable, but behavior change work is comparatively popular with the autistic and IDD communities (32.3% and 35.7% of papers, respectively). We thus encourage accessibility researchers to reflect on what research problems they address, and whether those problems are indeed priorities for the communities they want to serve. Increasing representation of disabled researchers, as emphasized by Spiel *et al.* [141], can help further identify community needs and research foci, while reviewing literature in fields that elevate Deaf and disabled people's voices (*e.g.,* Deaf studies and disability studies) is also critical.

Finally, a few other key areas of momentum are worth noting. As mentioned in Section 2 and related to the previous point, accessibility researchers are increasingly drawing on disability studies to inform their work (*e.g.,* [16,100,141]). Additionally, with the increase of data-driven AI systems and accompanying ethical issues, as emphasized by a recent ASSETS workshop [148], we call for more "dataset" contributions that are rooted in accessibility knowledge—a relatively sparse contribution type in our survey. Finally, contributions on best practices for teaching accessibility, as explicitly introduced for the ASSETS conference in 2019, are needed to help grow capacity and expertise within industry.

**5.2 Engaging with Users in Accessibility Work**

Compared to HCI as a whole, accessibility research introduces unique questions about how to engage with disabled participants and what role nondisabled participants, including specialists and caregivers, should have.

Accessibility research frequently includes disabled participants or older adults, but concerns can arise in the extent to which study procedures are accessible to these participants and the depth of their engagement. Regarding accessible procedures, the use of study locations that are accessible and/or convenient to participants, such as homes or community centers, is a positive research community norm. For depth of engagement, a relatively small number of papers employed participatory design (10% of user-study papers), a potentially powerful method to empower end users in design. However, some papers that used participatory design terminology included only single sessions with participants, reflecting recent critiques on what constitutes participatory design [13]. Moreover, there are unique challenges in an accessibility context, such as the potential overreliance and underacknowledged use of people with disabilities for their "access labor" in PD [17] and the inaccessibility of some PD methods to certain communities, such as people with aphasia [73]. We suggest that authors think carefully about these tradeoffs and that they discuss the tensions in their publications.

---

[9] We recorded people with multiple disabilities as "other" in our dataset rather than selecting two communities of focus categories since 1) they occurred very infrequently, and 2) the experience of a deaf-blind individual is not the sum of the experiences of a deaf individual and a blind individual.

Another persistent challenge in accessibility work is how to recruit sufficiently large samples for user studies [132]. We found a median sample size of 13 (*SD*=478.8) for disabled and older adult participant groups, but some specific communities of focus had much smaller samples (*e.g.,* samples of autistic people had a median of 9). The median of 13 is also lower than a recent analysis of CHI 2020 papers in general, which found lab, field, and interview studies had a median of 15-16 participants and remote studies had 182 [80]. We argue, however, that these small sample sizes are not necessarily a limitation, especially given the potential for accessibility researchers to burden participants by repeatedly sampling from small populations and the difficulty of participation for some people (*e.g.,* those with ALS) [17]. Methods that support smaller sample sizes (*e.g.,* case studies, single-subject experiments) or less interaction with participants can address these problems [132], though these methods were rare in our dataset: only 4% of papers included case studies with participants and 6% employed methods that did not require user studies. Content analyses and online ethnographies of user-generated information such as social media posts, videos, or other online content (*e.g.,* [56,72,122]) are particularly promising for supporting insight while requiring less direct participant engagement. We encourage the community to carefully consider broader HCI and methodological norms and how they apply to accessibility research, where it is difficult to recruit large numbers of participants. Moreover, we encourage reviewers to engage in discussions with the authors and program committee before dismissing an accessibility paper based on a small sample size.

Finally, our analysis reveals the roles of nondisabled participants in accessibility work. Two of these roles—ability-based comparisons and proxy participation—can sometimes be useful but must be approached with caution due to the risk of reinforcing normative, ableist beliefs. Comparisons can highlight divergent experiences as evidence in calling for a more just, accessible world, but when nondisabled people are conceptualized as the baseline "normal" there is also the danger of perpetuating rather than dismantling ableism. Disability studies has long challenged frameworks that place "normal" as the goal for which disabled people should strive [37]. The use of proxy participants raises similar ethical dilemmas. For both ability-based comparisons and proxies, we recommend that researchers first question their own motivation: is the proxy usage justified or is it simply easier than alternatives, such as learning how to alter their communication style to better match the participant's style (as demonstrated by Spiel *et al.* [191])? For proxies, another strategy is to engage with both disabled participants *and* proxies to triangulate the data collected from both (*e.g.,* [138]) and/or to allow the disabled participants to confirm or correct information from the proxies (*e.g.,* [74]). Regardless, for both comparisons and proxy use, we call on researchers to justify their decisions about these issues, which are often complex and nuanced, so that reviewers and other readers can effectively assess and critique the work.

### 5.3 Reflections on Our Research Process and Limitations

Rigorously curating a literature survey dataset and applying codes was challenging and effortful, and we would be skeptical of literature survey papers that do *not* discuss these details. The ACM provided us with a metadata dump of the CHI and ASSETS proceedings within our target time period, but cleaning that data took substantial manual effort to ensure the number of papers closely matched expected counts. Further, following our codebook development process that included IRR, we manually and programmatically looked for inconsistencies in the coded data and spent hours discussing difficult papers. In sum, these challenges have implications for how to assess future literature reviews and the importance for researchers to provide sufficient methodological detail. The data cleaning challenges are likely to be magnified when including publications from multiple sources (*e.g.,* ACM, IEEE, Elsevier, Thomson Reuters as in [19]) and relying on purely automated analyses (*e.g.,* [14,63,93]). And, while IRR may not be appropriate for reviews using a critical discourse analysis or other thematic approach (*e.g.,* [140]), it is key for more quantitative reviews.

In conducting this work, we also encountered highly varied language use across papers, which impacted our codebook and the conclusions we can draw from the work. Expanding on a similar conclusion by Brulé *et al.*'s [29] research with BLV users, blind or low vision people were referred to as "blind", "visually impaired", "low vision", "differently sighted", or having "vision loss". More difficult for our coding was the inconsistency in how papers described neurodiversity, and cognitive and/or learning disabilities. People with autism, IDD, or cognitive impairments were often grouped together in participant descriptions, for example, saying "special education students" or "psychosocial disabilities" (*e.g.,* [89,127]). In assigning these codes, we yielded to authors' descriptions of the work (*e.g.,* using the term "cognitive impairment" or "developmental disability"), and coded more general or unclear descriptions as "other". We acknowledge that other researchers may have derived different code categories, perhaps more coarse- or fine-grained ones, to describe neurodiversity and cognition-related disabilities.

Finally, beyond these challenges, our work has other limitations. First, our approach to filtering CHI papers and the codebook we derived are shaped by our positionality, and other researchers may have defined these steps differently. Second, though necessary for our quantitative study method, concretizing boundaries of what qualifies as a certain disability and what does not is itself somewhat problematic. The lines between communities are often overlapping or ill-defined; for example, someone who has an IDD or is a stroke survivor may encounter cognitive accessibility issues, physical accessibility issues, or both, and these distinctions are not always clear nor relevant in work focusing on functional differences rather than identity. Third, we scoped to only two popular conferences for accessibility work (CHI and ASSETS) so that we could conduct in-depth analyses. The norms and contributions (e.g., populations studied, sample sizes) identified in our study do not necessarily generalize to research published at other venues, including other HCI conferences and journals such as TOCHI and TACCESS.

## 6 CONCLUSION

In conclusion, our paper reflects on the field of accessibility's growth and history for the past 26 years, its implications for the field's current work, and where the community should go from here. Through a combination of qualitative and quantitative analysis strategies, we contribute both a deep understanding of work in the past decade and a broader overview of changes in the community for the past 26 years. While our results and discussion highlight a variety of beneficial practices already adopted in the community, we also suggest areas for improvement. For example, we encourage better representation in the community and promote the need to be cognizant of how ableism can surreptitiously be incorporated into research norms. We hope that researchers utilize our data and ideas to reflect and discuss with others the future of accessibility research.

## 7 ACKNOWLEDGMENTS

We thank Rose Guttman and Emily Rosenfield for their early help in data analysis and thank the ACM for supplying the data for our study. This work was funded in part by the National Science Foundation under grants IIS-1818594 and IIS-1652339 as well as the UW Center for Research and Education on Accessible Technology and Experiences (CREATE).